# Perspectives of Visualization Onboarding and Guidance in VA

Christina Stoiber, Davide Ceneda, Markus Wagner, Victor Schetinger, Theresia Gschwandtner,
Marc Streit, Silvia Miksch, and Wolfgang Aigner

**Abstract**—A typical problem in Visual Analytics (VA) is that users are highly trained experts in their application domains, but have mostly no experience in using VA systems. Thus, users often have difficulties interpreting and working with visual representations. To overcome these problems, user assistance can be incorporated into VA systems to guide experts through the analysis while closing their knowledge gaps. Different types of user assistance can be applied to extend the power of VA, enhance the user's experience, and broaden the audience for VA. Although different approaches to visualization onboarding and guidance in VA already exist, there is a lack of research on how to design and integrate them in effective and efficient ways. Therefore, we aim at putting together the pieces of the mosaic to form a coherent whole. Based on the Knowledge-Assisted Visual Analytics model, we contribute a conceptual model of user assistance for VA by integrating the process of visualization onboarding and guidance as the two main approaches in this direction. As a result, we clarify and discuss the commonalities and differences between visualization onboarding and guidance, and discuss how they benefit from the integration of knowledge extraction and exploration. Finally, we discuss our descriptive model by applying it to VA tools integrating visualization onboarding and guidance, and showing how they should be utilized in different phases of the analysis in order to be effective and accepted by the user.

✦

## 1 INTRODUCTION

The practical uptake of Visual Analytics (VA) approaches has increased substantially in recent years. Similarly to what happens in any data analysis process—and VA is no exception in this respect—analysts using VA typically alternate between a data *foraging loop*, in which bits of information are collected from a larger data pool, and a *sensemaking loop*, in which hypotheses are generated and explored by making sense of the data collected in the aforementioned process [49].

However, Pirolli and Card [49], who first described the sensemaking loop, mention that this process is not free from challenges: a set of *pain points*—or leverage points—typically affect the analysis and add to its total cost if no solution is provided to alleviate them. For instance, collecting information comes at the cost of exploring sufficient portions of the data and follow-up searches. Making sense of the data is limited by the human working memory and the user's attention span.

The literature published in the last decade can be seen as a generalized attempt to mitigate such problems. Improvements in the visual representations and the developments of modern visual interfaces are all meant to reduce the costs associated with data exploration and augmenting the human working memory. The advancements in the field of machine learning have led to a plethora of "smart-agents" that aim to expand the options available to the users, allowing them to "discover the unexpected" while cutting down the effects of possible biases. Knowledge can also be reused and exploited to reduce the efforts of analyzing domain-specific problems, in which the threshold to gain new insights is usually set higher due to the initial knowledge required to analyze the data in the first place.

Transforming data into insights and knowledge is a challenging and time-consuming process. What we can observe, though, is that the success of any data analysis process relies on a small set of ingredients: (1) a well-designed user interface supporting effortless data exploration, (2) appropriate visual data encoding to account for the cognitive capabilities of the human visual system and promote, in the best way possible, the completion of tasks, and (3) a sufficient amount of knowledge, either in the user's mind or externalized in a knowledge base, to make sense of the data.

These three ingredients, however, are often insufficient for successfully concluding the analysis. When the knowledge of the user does not match the expertise required to perform the analysis tasks, there is a knowledge gap that prevents the completion of the task. In recent times, smart approaches, namely visualization onboarding and guidance techniques, have been studied to account for such situations and to address different *knowledge gaps*.

Visualization onboarding and guidance are concepts easily understood, as they are also part of our daily experience. Imagine a cooking recipe with a list of ingredients but no instructions. Depending on the complexity of the dish, you might assume that flour, water, and eggs should be mixed to prepare the dough, however, this might lead to mistakes. Cooking instructions are a good analogy for guidance, clarifying what to do with the ingredients in order to reach the expected result (the cake). Anyone with some cooking experience knows that even with detailed instructions, recipes are not always easily reproducible. How can the cook know, for instance, that the "light brown" color of the bread in the oven matches the one expected in the recipe? Thus, onboarding is similar to detailed instructions and visual aids that are used for recipes. When a person is able to follow these instructions and reproduce the dish, the onboarding was successful.

Starting from this example, we argue in this paper that visualization onboarding and guidance are complementary and decisive dimensions of any successful VA approach. While visualization onboarding and guidance have received a fair share of attention in their own right, they have not been considered together, nor has it been described how they concur to help the user complete any given task. In this paper, we go in this precise direction, showing their different uses and scopes. We build on well-known models to pinpoint the subtle (but important) differences that characterize visualization onboarding and guidance across domains, describe the scenarios in which it makes more sense to use them, and show how they should be utilized for being effective and accepted by the user (Section 4). In Section 5, we discuss a usage scenario in which we apply our framework to the financial domain, using Why/Who/When/Where/How questions and showing how onboarding and guidance can facilitate two common tasks in

- *Christina Stoiber, Markus Wagner, and Wolfgang Aigner are with the St. Pölten University of Applied Sciences, Austria. E-mail: {name.surname}@fhstp.ac.at*
- *Davide Ceneda, Victor Schetinger and Silvia Miksch are with TU Wien, Austria. E-mail: {name.surname}@tuwien.ac.at*
- *Theresia Gschwandtner is with Group Central Compliance and Strategy, Erste Group Bank AG, Austria. E-mail: Theresia.Gschwandtner@erstegroup.com*
- *Marc Streit is with Johannes Kepler University Linz, Austria. E-mail: marc.streit@jku.at*

financial analysis: building a portfolio, and swing trading. The object of our analysis is the software *Profit* [1]. While we do not provide information regarding design rationales for visualization onboarding and guidance (as discussed in our previous work [11, 13, 61]), our overall goal is to support the understanding and the intertwining of these two concepts to foster a well-considered usage of visualization onboarding and guidance techniques in practical scenarios, thereby reducing the cost of making sense of data, as initially promoted by Pirolli and Card [49]. In summary:

- We present a descriptive model of how visualization onboarding and guidance can be effectively integrated into the VA process (Section 3).

- We confront the two concepts of visualization onboarding and guidance and analyze their suitability in different analysis settings using Why/Who/When/Where/How questions (Section 4).

- We describe a usage scenario of our model using the stock trading VA tool *Profit* to show how visualization onboarding and guidance can support users in different phases of the analysis (Section 5).

## 2 RELATED WORK

In this section, we revisit the literature addressing the two topics of visualization onboarding and guidance. Both of them typically also rely on the presence of a knowledge base. Therefore, we complement our discussion by presenting knowledge-assisted VA as well.

### 2.1 Visualization Onboarding

User Onboarding in HCI: The term "onboarding" originates from organizational theory where it is widely used and refers to the step taken by an organization to facilitate and socialize newcomer adjustment [37].

The topic of onboarding has received attention by the user experience (UX) practitioner community. Several blogs and articles present guidelines and design inspirations for onboarding concepts [4, 24, 25, 27]. UX practitioner Hulik [67] defined *user onboarding* as "the process of increasing the likelihood that new users become successful when adopting your product." [38]. Examples of onboarding design patterns include: instructional text, tours, progress bars, just-in-time hints, tips placed in feeds of user-generated content, and interactive tutorials [4]. Hulik and Higgins [24, 25, 27] provide a set of guidelines and processes to engage with the challenge of onboarding design and integration.

Towards Visualization Onboarding: The process of mapping data variables to visual channels is the central component of virtually all known conceptual models of visualization, such as [10, 68]. Understanding the visual mapping is key for the correct decoding of both the visual representation and the underlying information. The other steps of the visualization pipeline—including data analysis, filtering, and rendering—also influence the appearance of the visual representation and need to be transparent and understandable for the user. Particularly for visualization-illiterate users in data analysis and visualization, however, this task is often difficult and prone to the risk of drawing the wrong conclusions or insights regarding the data [6, 7, 9]. Visualization onboarding concepts can help users comprehend the visualization process and support learning. Stoiber et al. [61] define visualization onboarding as follows: *Visualization onboarding is the process of supporting users in reading, interpreting, and extracting information from visual representations of data.*

Visualization Onboarding Approaches: So far, there has been little discussion about onboarding concepts for visualization techniques and VA tools. Tanahashi et al. [64] investigated top-down and bottom-up teaching methods as well as active and passive learning types for scatterplot, graph, storyline, and treemap representations. The bottom-up teaching method focuses on small, detailed pieces of information which students then put together to reach comprehensive understanding. A top-down teaching method starts by presenting broad overviews to the students, thus helping them understand the abstract, high-level parts of an idea which then provides the context for understanding its components in detail [64]. Passive learning means that students only receive the information, without participatory dialog. In contrast, active learning describes an active participation. Their analysis indicates that top-down methods were more effective than bottom-up methods. Additionally, their study showed that utilizing the active learning type with top-down tasks proved to be the most effective. Kwon and Lee [39] explored the effectiveness of active learning strategies. Three tutorial types—static, video-based, and interactive—are used to support the learning of scatterplot visualizations. They observed that participants using the interactive and video tutorials outperformed participants with static or no tutorials. Ruchikachorn and Mueller [53] proposed a concept of teaching by analogy, i.e. demonstrating an unfamiliar visualization method by linking it to a more familiar one (see Figure 6). They provide demonstrations of various visualization techniques, e.g., data tables and parallel coordinates; scatterplot matrix and hyperbox; linear and spiral charts; hierarchical charts and treemaps. Yalçin [74] presented the design of a contextual in-situ help system for the visual data interface Keshif [36], called HelpIn, to explain the features of this tool. Recently, Wang et al. [72] presented a set of cheat sheets to support literacy around visualization techniques inspired by infographics, data comics, and cheat sheets in other domains.

Nowadays, most commercial visualization tools already integrate some basic onboarding concept focusing on the explanation of features. *IBM Cognos Analytics* [29], for example, uses step-by-step tours with tooltips and overlays for onboarding new users (see Figure 5). The commercial visualization tool *Advizor* [30] uses more traditional textual descriptions to explain the visual mapping for various visualization techniques.

### 2.2 Guidance

Scientific work related to guidance goes back to the development of the first human-computer interfaces [16, 17, 60]. In this initial phase, guidance was mentioned only as an abstract guidelines for an effective design of user interfaces. The term guidance has been used also in other domains. For instance, in decision theory Mark Silver described how to provide assistance through decision support systems [59].

Although using many different names, a considerable number of guidance approaches have been developed even in the visualization community. A preliminary work by Schulz et al., aimed at bringing all such approaches under the common name of guidance and described an initial set of characteristics they all had in common [58]. Among many others, one renowned example of guidance are the Scented Widgets [73]. The approach uses visual cues integrated into interface widgets for guiding the user through the exploration of a large information space. Gotz et al. [20] describe how to assist users in creating their own visualizations and solve specific visual analysis tasks. The approach automatically understands the user's task by matching the interaction patterns with a database of previously created visualizations and accordingly proposes improvements to the current design. More recently, Ceneda et al. [13] describe guidance to support the analysis and exploration of cyclical patterns in time-oriented data. In particular, the VA tool they designed is able to provide the user with indications about the most prominent cycles to explore, the length of the cycles and possible repetitions in time. More recently, a literature review by the same authors categorized these and many other guidance approaches focusing on the mixed-initiative nature of guidance methods [12]. Guidance can be, in fact, seen as a dual process in which not only the system supports the user's analysis, but also in which the user supports and provide valuable feedback to the analytical system and steer the analysis.

Although there has been much progress in the field, there is still a lot of confusion about how to distinguish approaches that qualify as actual guidance from those that do not. When it comes to guidance in VA, the confusion might also originate from a vague definition of the types of knowledge gaps that can be addressed by means of guidance [11]. For this reason, we propose a unified model, clarifying the extent of visualization onboarding and guidance in VA.

## 2.3 Knowledge in Visual Analytics

The discovery, acquisition, and generation of new knowledge are the main goals of VA. According to Thomas and Cook [65, p. 42], the final task of the analytical reasoning process is to create some kind of knowledge product or direct action based on gained insights. However, this process is not always straightforward. When approaching data analysis, we can imagine that the user has a certain amount of *prior* knowledge [14] that is based on previous experiences. We refer to this as *tacit* knowledge [19, 71], as it is usually gained through incidental activities and is mainly contained in the user's mind. However, as described in Section 5, in increasingly complex data analysis scenarios, this knowledge is not sufficient to fully exploit the features of the analysis tools and, as a consequence, to solve all tasks and complete the analysis. This problem stems from a mismatch between the knowledge required to perform the task and the knowledge possessed by the analyst. We refer to this issue as the *knowledge gap* [11, 19, 61]. Therefore, user assistance is intrinsically related to the generation of knowledge, in that it actively contributes to the closing of knowledge gaps.

The process of closing a knowledge gap through visualization onboarding and guidance greatly benefits from the use of external sources of knowledge. As previously mentioned, a certain amount of knowledge is typically required to operate tools, namely *operational* knowledge (how to interact with the visualization system), and *domain* knowledge (how to interpret the content) [14]. This knowledge can be externalized, stored in a database, and actively used to support the user when a knowledge gap is detected [19, 52, 71]. We call this process *externalization*, and its output is *explicit* knowledge [19, 71]. Stoiber et al. [61] further enhanced these two levels of knowledge: (1) *domain* (e.g., vocabulary and concepts); (2) *data* (understanding the particular data type); (3) *visual encoding* (understanding the visual mapping); (4) *interaction concepts* (for performing tasks and understanding relationships between the data); and 5) *analytical knowledge* (knowledge of different automated data analysis methods). *Operational knowledge* [14] can be seen as a combination of *visual encoding, interaction, and analytical knowledge* [61].

We will use the aforementioned different types of knowledge to show how visualization onboarding and guidance deal with them.

Domain Knowledge: This kind of knowledge is related to a specific field of interest (e.g., healthcare, journalism, cyber security). Each domain has its own vocabulary [47] for describing the data, workflows, conventions, and how data can be used to solve a problem [46]. Domain knowledge is also an ensemble of concepts, intellectual tools, and information resources that a user can draw upon to put the visualized data into context. An appropriate solution to a lack of domain knowledge is a knowledge base containing an externalized and computer-readable version of such knowledge (i.e. explicit knowledge $\boxed{\mathcal{K}}$) [19, 70, 71]. One example would be a collection of computer-executable medical guidelines. On the one hand, onboarding techniques can use this knowledge to support the user's contextual understanding. For instance, by describing the vocabulary used, the domain's conventions, workflows, and explaining why certain choices should be / have been made. Onboarding might provide information as to why a particular treatment plan was selected based on the results of a blood test. On the other hand, the knowledge base can be exploited for providing guidance, for instance, to support the discovery of insights, or to decide on future steps, such as choosing the next medical treatment [18]. Guidance might provide information on which treatment method promises the greatest success for a certain evaluation result of a blood test in combination with the patient's other key data.

Data Knowledge: The first step in any VA process is data selection and manipulation [35]. Therefore, when approaching a problem, the user must first possess knowledge about the data at hand. Onboarding techniques tackle a lack of low-level information about the data, such as understanding principles of the specific data format, data types, or data structure. In contrast, guidance leads the user through the VA process for manipulating a specific data set as well as showing or highlighting interesting aspects of the data, such as outliers, missing data points, uncertainties, inconsistencies.

Visualization and Interaction Knowledge: This type of knowledge deals with understanding the visual mapping [10] and the interaction with the VA environment. By avoiding confusion about how to use the provided interaction features, or which effects these interactions have, onboarding as well as guidance can fill a knowledge gap regarding the visualization or the interactions.

When dealing with interactive visual representations, onboarding mainly addresses problems related to understanding the visual encoding and the interaction concepts, performing tasks, and understanding relationships within the data. Typically, onboarding methods aim to fill this knowledge gap by using contextual menus, labels, or tutorials, so that the user can understand their use and function [61]. In contrast, guidance deals mainly with formalizing and structuring execution plans using these controls and interaction means as building blocks. Thus, guidance might be used to help the user gain new insights from the visualized data, e.g., by highlighting deviations from normal medical conditions in the blood test data indicating a specific disease.

Analytical Knowledge: This knowledge is necessary to understand the analytical methods that are used by the given VA approach and how to set the parameters (e.g., machine learning techniques for prediction or classification). The user needs to have a basic understanding of the analytical methods and their characteristics in order to be able to effectively choose, parameterize, or utilize them. This is where onboarding techniques come into play [61]. They establish a basic understanding and introduce the user to the different possibilities of the analytical methods. Furthermore, assisting the user in choosing the most appropriate of these analytical methods and parameters with respect to the task at hand [8] is a typical usage scenario for guidance. Sometimes, this might also include the comparison of performance metrics or an estimation of user preferences.

In this work, we describe how user assistance can benefit from explicit knowledge sources and contribute to the generation of new knowledge and insights. Based on the previously introduced terminology, we further characterize knowledge in Section 4 by listing all possible gaps that might arise during analysis and show how user assistance can contribute to their resolution.

## 2.4 Model Building

Up to now, a number of papers have presented different methods dealing with the integration of knowledge in the VA process. Thereby, we can distinguish between descriptive models and mathematical models. In the view of descriptive models, Lammarsch et al. [41] described that the combination of automated analysis methods with interactive visualizations is "a necessary step". The framework by Sacha et al. [55] describes the process of knowledge generation based on three dedicated loops (evaluation loop, verification loop and knowledge generation loop) when a human analyst is using VA tools. Furthermore, Sacha et al. [54] have illustrated how uncertainties arise, propagate and impact human knowledge generation processes by relating the concepts of uncertainty, awareness, and trust. A recent paper by Andrienko et al. [3] presents a framework in which the VA process is considered as a goal-oriented workflow producing a model as a result. Besides, Chen et al. [15] proposed a general framework bridging the gap between VA and storytelling. They introduced a story synthesis phase that extends the VA workflow with storytelling.

Van Wijk [69] presents a generic visualization model discussing the cost and gains of VA as well as the integration of knowledge on the users side of the model based on mathematical equations. As extension, Wang et al. [71] described in their paper that the integration of knowledge into the visualization process for solving analytical tasks is a fast growing area. By integrating the experts knowledge into the visualization process, "the experts are more capable of performing complex analytical processes" [71, p. 616].

The presented descriptive model (see Section 3) is also conceptually grounded in the visualization model introduced by Van Wijk [69] as well as on the "Knowledge-Assisted Visual Analytics Model" by Federico & Wagner et al. [19, 70].

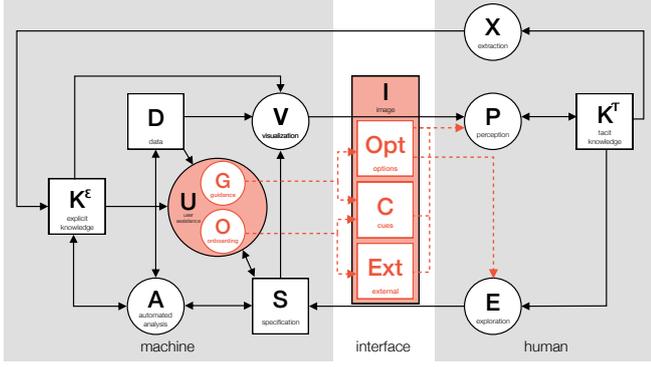

Fig. 1. The overview model of user assistance (drawn in black) shows the incorporation of user assistance $(U)$ into the "Knowledge-Assisted Visual Analytics Model" [19] with connections to data $\boxed{D}$, explicit knowledge $\boxed{K^\varepsilon}$, and specification $\boxed{S}$. The detailed model representation (drawn in orange) shows that visualization onboarding $(O)$ and guidance $(G)$ are contained in the user assistance $(U)$ process. Additionally, $(O)$ and $(G)$ are utilizing options $\boxed{Opt}$, cues $\boxed{C}$ or external resources $\boxed{Ext}$ contained in the image $\boxed{I}$ as well as the exploration $(E)$ and the perception $(P)$ shown by dotted orange lines.

Summarizing, we discussed the work related to the three main topics: visualization onboarding [61], guidance [11] and knowledge-assisted VA [19]. Onboarding can support users in learning, and correctly interpreting applied VA methods [61], whereas the intuitive or pervasive nature [21] of guidance [11] supports users throughout the visual exploration. Additionally, with knowledge-assisted VA [19], implicit knowledge of the user can be extracted and provided as machine-readable knowledge used for various aspects of onboarding and guidance. In the next section, we present the descriptive model in detail.

## 3 DESCRIPTIVE MODEL OF INTEGRATING GUIDANCE AND VISUALIZATION ONBOARDING

To promote a discussion of visualization onboarding and guidance analogies and differences and to foster a better design and integration of such concepts into the VA process, we present a general descriptive model [5] combining and describing the aspects of **visualization onboarding** and **guidance** under the umbrella term of what we might generically call **user assistance** (see Figure 1).

In order to illustrate the role of visualization onboarding and guidance in VA, we added the user assistance $(U)$ element—containing onboarding $(O)$ and guidance $(G)$—to the machine space of the model, as it is aimed at semi-automated generation rather than provided by a human expert (see Figure 1). For providing user assistance $(U)$, connections between extracted and stored explicit knowledge $\boxed{K^\varepsilon}$, data $\boxed{D}$, and the system's specification $\boxed{S}$ are needed. In particular:

1. *Data* (data tables) is used as input for user assistance ($\boxed{D} \rightarrow (U)$) (e.g., important meta-information, such as the dimensions of the data, the included data types, and the data structure).

2. *Explicit knowledge* is another potential input to user assistance ($\boxed{K^\varepsilon} \rightarrow (U)$). For instance, the knowledge gap of the current user might be identified and closed based on the stored explicit knowledge of other users. Thus, user assistance can be provided with respect to the state of knowledge.

3. *User assistance* and *specification* are connected in both directions ($(U) \leftrightarrow \boxed{S} \leftarrow (E)$) and the exploration state is an input to the specification. For example, the current exploration state defines the VA system's specification. If the user needs assistance, the current specification settings are used to determine appropriate user assistance. Based on the assistance steps, the user interacts with the system, thereby changing the specification.

Figure 1 shows the two user assistance $(U)$ concepts, integrating elements of guidance $(G)$ and onboarding $(O)$, that are highlighted in orange. Here, user assistance $(U)$ is directed from the machine side towards human perception $(P)$. The onboarding $(O)$ and guidance $(G)$ processes both exploit similar input information. Based on their goals, different types of assistance can be provided (see Section 4). In the following subsections, we discuss how to integrate visualization onboarding and guidance into the VA process.

### 3.1 The Integration of Visualization Onboarding
The detailed description of the onboarding process in combination with all other elements is illustrated in Figure 1:

$$\{ \boxed{K^\varepsilon}, \boxed{D}, \boxed{S} \} \rightarrow (O) \rightarrow \{ \boxed{C}, \boxed{Ext} \} \rightarrow (P)$$

Onboarding can be either directly integrated into the interface of the VA tool in the form of visual cues $\boxed{C}$, or independently as external source $\boxed{Ext}$—for example a website of the VA tool with instructional material. Visual cues $\boxed{C}$ for onboarding may be represented by tooltips, annotations, overlays, labels, highlighting in contextual menus, etc. [61] or, as presented in Figure 7, sticky notes. Additionally, external resources $\boxed{Ext}$—commonly websites or documents describing the VA tool—provide explanations based on screenshots, videos, textual descriptions, etc. (see Figure 4). In general, onboarding methods support the user in understanding different aspects of a VA system such as the visual encoding, interaction concepts provided for the exploration $(E)$, the underlying data $\boxed{D}$, or analytic methods $(A)$ [61]. Explicit knowledge $\boxed{K^\varepsilon}$ may act as the basis for adaptive onboarding methods, supporting the closing of the user's knowledge gap, for instance understanding the visual encoding of the presented visualization technique.

### 3.2 The Integration of Guidance
$$\{ \boxed{K^\varepsilon}, \boxed{D}, \boxed{S} \} \rightarrow (G) \rightarrow \{ \boxed{Opt}, \boxed{C} \} \rightarrow \{ (P), (E) \}$$

Different types of guidance might be advised for different users. For instance, visual cues $\boxed{C}$ could be provided that are either integrated directly into the visualization itself or as part of the analysis environment. Visual cues usually present a set of unordered suggestions on how to proceed with the analysis. Other visual cues might be useful for orientation purposes, i.e., showing users where they currently are. Through the perception process $(P)$ (i.e., perceiving these visual cues), they help the analyst to make sense of the data, and thus to gain insights. In addition, guidance could be communicated by providing the analyst with various options $\boxed{Opt}$ of how to proceed with the analysis. Usually, these options are ordered, for instance by importance, and help the analyst to navigate through the exploration process. Finally, guidance could also provide a step-by-step description of how to proceed. This kind of guidance autonomously influences the specification ($(G) \rightarrow \boxed{S}$) of the visualization and of the analytical processes. In contrast to guidance, onboarding does not provide any options giving concrete suggestions on how to proceed in a given situation.

## 4 ASPECTS OF VISUALIZATION ONBOARDING AND GUIDANCE
A common goal of visualization onboarding and guidance is to support the user in effectively utilizing a specific VA tool. Besides this common goal, however, they differ in the way they support it. While onboarding is targeted at teaching the user how to use the tool, guidance provides interactive assistance while using the tool. In the following, we present aspects of visualization onboarding and guidance based on five questions [22, 23]. Figure 2 gives an overview of the differences and similarities of visualization onboarding and guidance and illustrates how they complement each other.

**Why** do we provide visualization onboarding and guidance? Although the general goal of both concepts is clear, different types of support should be provided with respect to the context in which it is required. In fact, visualization onboarding and guidance serve different purposes. Users have difficulties interpreting and working with novel visual representations or comprehending characteristics of the underlying data. Therefore, onboarding mainly deals with providing hints to understand and get acquainted with the interactive visualization environment—often before the actual use of the tool—or to tackle a lack of general understanding during the first use. Moreover, guidance aims at assisting the user in overcoming knowledge gaps regarding how to find or reach a specific analysis goal during an interactive analysis session. Finally, onboarding mainly involves perception and interaction, whereas guidance encompasses cognition and reasoning for knowledge discovery. Both concepts can help to reduce the costs of working with an VA tool by providing assistance at different stages of the usage of the tool.

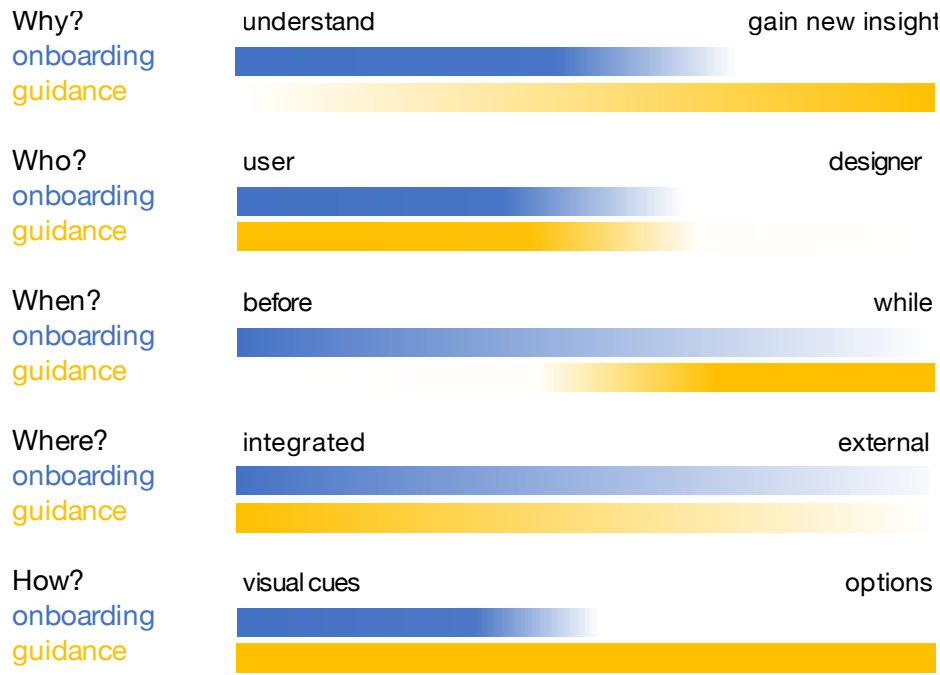

Fig. 2. Aspects of visualization onboarding and guidance along the questions of **Why** do we provide visualization onboarding and guidance?, **Who** can benefit from visualization onboarding and guidance?, **When** to provide visualization onboarding and guidance?, **Where** to provide visualization onboarding and guidance?, and **How** to provide visualization onboarding and guidance?

**Who** can benefit from visualization onboarding and guidance? Any user of the VA environment can benefit from visualization onboarding and guidance. In fact, the proper implementation of both concepts can be critical for a solution to support a wide variety of users. Individuals have different amounts of prior knowledge and therefore different *knowledge gaps* [11, 19, 61] when using a VA system to complete a certain task. Visualization onboarding and guidance do not target the designer of the VA systems. However, designers can benefit from the descriptive model, as they can use it as a blueprint (described in Section 6.2).

**When** to provide guidance and/or onboarding? Another important difference between visualization onboarding and guidance is the moment in which they are required (see Figure 2). Usually, onboarding means are offered immediately before or during the analysis, although many of them take place during the first use of a specific element, for instance an introductory tutorial [61]. Guidance, on the other hand, is provided mainly during the analysis. However, for guidance techniques, this questions can be tricky to answer since all analysis states are typically not known in advance. Guidance is especially beneficial at the moment when the user has to make a decision [59].

**Where** to provide visualization onboarding and guidance? This question aims at defining the locations where assistance should be offered—either onboarding and/or guidance. Typical locations for providing both visualization onboarding and guidance are found in the visualization environment itself [34, 62, 74] (i.e., integrated into the visualization or as a separate component). An example of internal onboarding can be seen in Figures 7 and 3. Kang et al. [34] presented approaches to help users get started with visualization interfaces using the sticky notes (see Figure 7) metaphor. Ymap is a map-based visualization tool where users can click on the map to see facts about the selected area in a table. Additionally, they can select multiple areas and zoom in on the map. A dynamic query to filter the map according to a list of criteria is integrated. A scatterplot that is coupled with the map and table shows relationships between these criteria. However, onboarding methods can also be offered externally [45, 48, 50, 63, 66]. For instance, common external onboarding resources are websites of the VA tool itself or offline documentations. A special case is represented by learning environments [2, 53, 64]. These can be small games that users may play before using an application in order to become familiar with the basic concepts or websites such as dataVizCatalogue [51] and FromDataToVis [26] which are libraries of different information visualisation types explaining the visual encoding. These learning environments cannot be considered as internal or external, as they are not directly related to the VA tool. For guidance, an external source would be represented by users providing assistance in real time during the analysis. However, it is usually preferred to avoid additional users as sources of guidance since they may introduce biases. When this happens, the necessary input for guidance should be adequately formalized and included in the knowledge base.

**How** to provide visualization onboarding and guidance? The last and most important question is related to the modalities for providing visualization onboarding and guidance. Answering this question correctly has a direct influence on the effectiveness of the provided assistance. However, the 'how' is context-sensitive and should therefore be dynamically generated, while also considering possible user preferences.

**Onboarding** procedures should be tailored to the users' preferences. Some might prefer videos or hands-on tutorials, while others might find textual descriptions or offline documents more useful [61]. An example how to integrate onboarding using visual cues can be seen in the approach of Kang et al. [34] using the sticky notes metaphor. These cues are context-sensitive and embed an onboarding method which supports the users by highlighting the main functions of the interface. The sticky notes provide a series of steps to accomplish a task, such as highlighting unconventional features or showing where to click in the interface). Furthermore, Yalcin [74] used overlays including a combination of *topic listing, point & learn, guided tour, notification, and topic answers* for his approach. Once the help material is selected, its answer is presented in situ, that is, the material is fully integrated into the interface. Relevant components are highlighted, tooltips to describe the actions are shown, and textual instructional descriptions are provided to the user (see Figure 3).

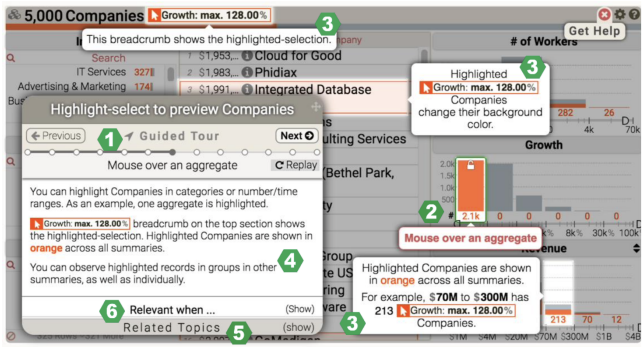

Fig. 3. The guided tour mode [74]: (1) The tour progress is visible, and users can direct it forwards, backwards, or to a specific step. (2) The tooltip of the main action, which is a mouse-over on a visual glyph, is highlighted by color. (3) Additional tooltips describe the effect of this action on other interface components. (4) A detailed description of the topic presents an easy-to-read summary of tooltips and additional information. Related topics and the context in which this topic applies can be viewed on demand as well.

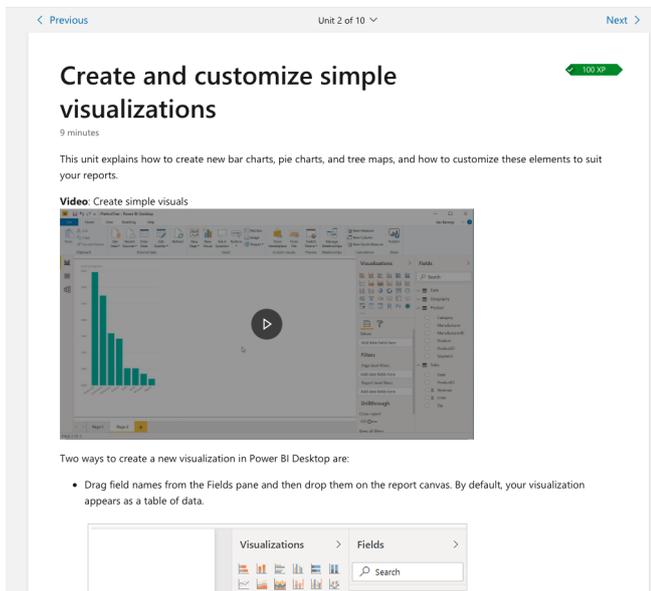

Fig. 4. External help website including videos, screenshots and instructional text to explain the features of Microsoft Power BI [45].

Commercial visualization tools in particular make use of documentation/explanation websites with screenshots, videos, and textual descriptions [50, 56, 57, 63, 66]. In Figure 4, the external help website of Microsoft Power BI [45] is presented including videos, textual descriptions and instructional texts.

The degree of **guidance** can also be varied (i.e., low to high level of guidance) and should be chosen with respect to the users' needs. Different degrees of guidance could range from visual cues for orientation to specific step-by-step instructions [11]. In general, using visual cues serves the goal of providing subtle suggestions to the user who, however, has the duty to discern and reason about them before blindly following suggestions. For instance, Luboschick et al. [42] use visual highlighting cues to indicate data cases that are worth analyzing based on a heterogeneity metric, and hence, foster profitable data exploration. Streit et al. [62] present possible future analysis paths to the analyst. Ceneda et al. [13] provide suggestions of potentially interesting cyclical patterns in time series data to users. Interesting cycle lengths are highlighted directly in the interface widgets that are used for exploring the data.

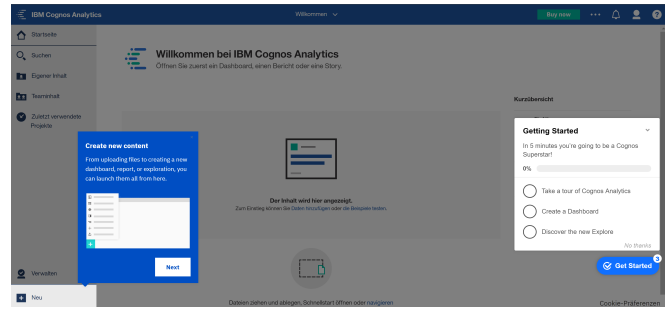

Fig. 5. Step-by-step tour in IBM Cognos Analytics [28] to onboard first-time users on demand using tooltips and overlays to explain the general features of the tool.

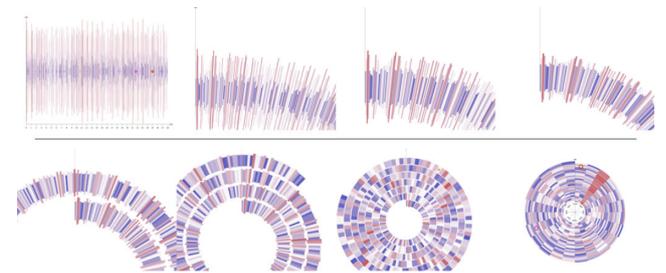

Fig. 6. Visualization onboarding by analogy. To introduce a spiral chart, a linear chart is gradually morphed into a spiral layout [53].

Guidance can also provide a set of alternative options for the next analysis steps. The user may or may not choose to follow these options. A typical example of this kind of assistance is recommender systems. The recommendations produced by such systems are directly related to the task the user is currently solving. In comparison to lower degrees of assistance, like visual cues, such assistance is considered stronger and directly tailored towards steering the course of the analysis. An example of this kind of assistance is presented by Kandel et al. [32, 33]. In their work, they designed Profiler and Wrangler, tools for supporting data profiling and transformation. The tool works by recommending a set of useful functions to perform data cleansing and transformation operations, based on the type of data that is analyzed. Another approach that aims to redirect the analysis is presented by May et al. [44]. Their approach aims at assisting the user in exploring large graphs by incorporating visual glyphs to direct the exploration towards interesting areas of the graph, or towards data that has not been analyzed yet.

One of the least intrusive forms of guidance is exemplified by approaches improving the user's orientation. In data analysis this typically translates into supporting the construction of a mental map, which refers to the ability of a user to build an internal representation of the data under analysis. Guidance works by directly targeting the user's perception exploiting low-level information extracted from the dataset, e.g., statistical values, and mapping such values to basic pre-attentive features, e.g., color hue, position, etc., drawing the user's attention towards certain data characteristics [12]. Such orientation cues could be added on top of the visualization itself or in the interface widgets. In both cases, the goal is to guide and address the interaction, foster understanding, and take the first steps towards gaining insights. The fact that orientation guidance works at such a low level means that it has to target basic human abilities. Typically, this involves the user's pre-attentive skills. For instance, highlighting data items or parts of the interface is typical of orientation guidance. The contrast between the color hue or the color intensity of the highlighted elements and the surrounding ones should immediately strike the user's attention, avoiding tedious search operations. May et al. [43] show an instance of such a technique: they provide guidance to feature selection for model building. This is achieved by changing the color hue, i.e., highlighting of data columns that the system deems interesting.

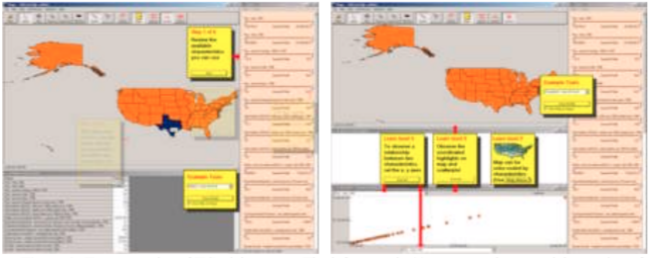

Fig. 7. Sticky notes [34] used to onboard users to the Ymap tool. This is an example of using visual cues to support users in understanding the main functions of the interface.

While context-free visualization onboarding and guidance are theoretically possible, they should be avoided as unspecific assistance might be even worse than doing nothing. In addition, the combination of multiple types of assistance might be indicated. For instance, a visualization tool could offer visualization onboarding to facilitate the use of the visual interface, while at the same time providing guidance options to complete a given task.

## 5 USAGE SCENARIO ON VISUALIZATION ONBOARDING AND GUIDANCE IN THE FINANCIAL DOMAIN

In this section, we describe how visualization onboarding and guidance can be effectively utilized to actively support domain tasks. To the best of our knowledge, not many approaches in the literature describe how to exploit and combine the full extent of visualization onboarding and guidance support. However, we were able to identify multiple aspects of visualization onboarding and guidance in the professional trading and stock market analysis software *Profit* (see Figure 8). Therefore, we provide conclusions regarding the financial domain from our descriptive model perspective along the Why/Who/When/Where/How questions in Section 5.1, as we did it in Section 4 to show the general aspects of visualization onboarding and guidance.

The financial domain is complex and hard to fully grasp, both in theory and practice. It is riddled with jargon and tangles legal, economic, and political aspects that can be both national and/or international. The state of the art in investing is also never truly reflected in the published literature, for obvious reasons. Due to its competitive nature, it is not in anyone's best interest to reveal optimal practices because they might provide economic advantages. Furthermore, it is commonly accepted that there is a division between the types of stakeholders that make up the economy (also known as "Wall Street vs. Main Street" [40]), which can cause wildly varying pictures of what good investments are at any given time. In this section, we show how visualization onboarding and guidance are essential for countering this complexity and providing efficient VA for the financial domain.

In general, financial analysis is, broadly speaking, divided into Fundamental Analysis (FA) and Technical Analysis (TA). FA places emphasis on *fundamentals*, qualities or quantities that can be measured about a stock, such as a company's equity, its brand value, or the skill of the executive team. To figure out a good investment strategy, investors can look at all data available to better gauge values. TA, on the other hand, assumes that the current price approximates the real value of a stock, all things considered. Its modus operandi is the visual analysis of graphs. It does not try to avoid overpricing, speculation, random events, and emotion as noise over the price signal. Rather, finding good investments involves using it tactically. Therefore, it is a useful analogy to think that FA is more of a strategic tool, while TA is more tactical: FA believes that fundamentals are the true latent variables that control the price over time, therefore it plans to win a war. TA believes that the stock signals have local behavior that can be predicted over shorter periods of time, allowing the user to corner the price and generate profit over many trade battles. These two approaches are not exclusive, and most investors will have a good understanding of both, while favoring one of them.

All trading tasks involve analyzing price behavior: inspecting the past, gauging the present, and estimating the future. Figure 13 shows three optimal trading patterns over the same candlestick series, all of which involve buying and selling stocks and profiting from the price difference, but with different strategies or restrictions. From images a to c, the amount of actions performed increases, the estimated profit increases, but so does the risk involved. These optimal patterns would not be achievable in a normal scenario, as it is impossible to determine the exact behavior of the price series.

The object of our analysis is the software *Profit* [1]. It is the largest platform for stock trading in Brazil, with over six million trades daily. This is a considerable share of all BOVESPA (Brazil's stock exchange) transactions at any given day. *Profit's target audience is personal traders* instead of large institutions and hedge funds, which represent the bulk of the financial volume. This makes it interesting for our exposition, as it is tailored to ease novice users in the financial domain into trading, which is reflected in the visualization onboarding procedures. Figure 8 shows *Profit's* main screen when opened for the first time. This is the first dashboard presented to new users but it is highly customizable, with hundreds of choices between different interfaces, indicators, buying interfaces, and views. With all the different parameters and possible financial instruments to be displayed, the possible dashboard configurations are virtually infinite. The main screen, illustrated in Figure 8 shows the valuation of the Brazilian stock exchange (BOVESPA) through the IBOV index in a candlestick chart. The reader might notice that it has the appearance of an analysis tool for temporal data, but the complexity of the domain can already be grasped by the amount of information displayed, and encodings which might be unfamiliar, such as the candlestick chart used for displaying price variation (Figure 12). Therefore, there is a need for visualization onboarding and guidance to support the users in exploration tasks with the *Profit* tool. To illustrate our discussion, we propose two trading tasks that can be predominantly associated with FA and TA, respectively: building a portfolio, and swing trading. Figures 9 and 10 show two desktops configured to perform the respective task within *Profit*. In the first example, a stock is bought and held for a long period of time, to be sold for a small but safe margin of profit (Fig. 13 (a)). The average value of a stock's price tends to increase slowly over time, so holding will probably result in gains even against local lows and fluctuations (i.e., the two drops in price between buying and selling). Swing trading, on the other hand, tries to use these fluctuations to benefit on many "buy low, sell high" trades over time, having a higher profit margin potential (Fig. 13 (b)). Both personalized desktops shown in Figure 9 and 10 subtly reflect these particularities of the task. In the first one, most of the screen space is dedicated to identifying good candidate stocks, and there is no button for buying stocks visible, as this action is done minimally and can be performed through a pop-up window. On the right, however, an integrated component for configuring and placing orders is strategically placed next to the graph.

### 5.1 Descriptive Model illustrated in the Financial Domain

In this section, we tie the two example tasks from FA and TA to our Why/Who/When/Where/How questions and provide conclusions re-garding the financial domain from our framework's perspective. In Figure 11, we present the aspect of visualization onboarding and guidance compared to the two tasks of FA and TA, according to the aspects described in Section 4.

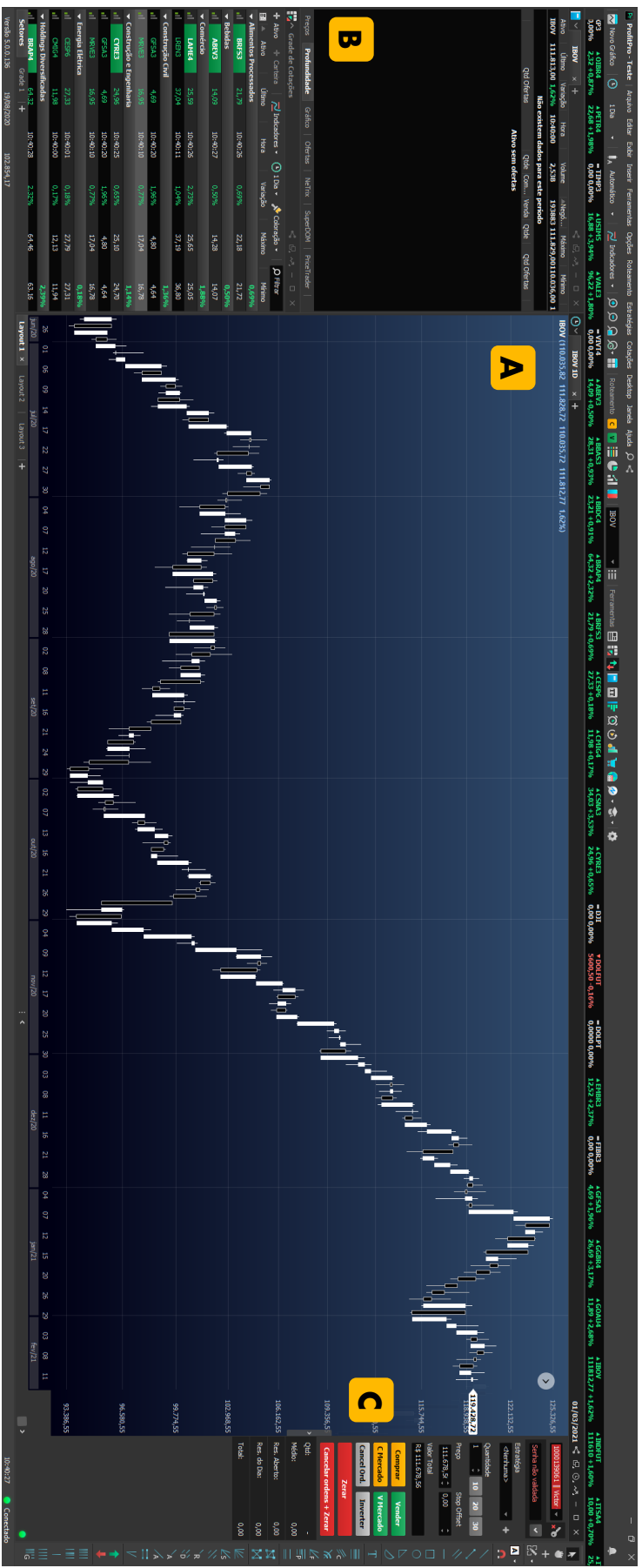

Fig. 8. Starting screen of the *Profit* [1] software in the default PT-BR language. The central window A shows the valuation of the Brazilian stock exchange (BOVESPA) through the IBOV index in a candlestick chart. Around it, a myriad of green and white numbers display other stock variations B. On the right C, controls for placing others and adding markers to the graph can be seen. This is the first dashboard presented to new users but it is highly customizable, with hundreds of choices between different interfaces, indicators, buying interfaces, and views. With all the different parameters and possible financial instruments to be displayed, the possible dashboard configurations are virtually infinite.

**Why** do we provide visualization onboarding and guidance? When evaluating the financial domain from a VA perspective, such as understanding the data, users, and tasks [46], two core aspects can be highlighted. First, it essentially uses time-series data, and second, most tasks have a standard key performance indicator: the profit. All the diversity of investing and trading strategies will fall under the scope of this umbrella, but they do not fully characterize what can be considered data analysis in the financial domain. A data journalist who writes for a financial newspaper can perform the same tasks for analysis as an investor, but only to gather information for publication, without the commitment of trading and therefore without the risk. This is an important distinction because it is evident that both visualization onboarding and guidance for these two users (the data journalist and the investor) in an *ideal scenario* would be similar, but cannot be the same.

When a task involves taking an action with risk, this should be reflected both in visualization onboarding (e.g., guaranteeing that the risk and its factors are clearly communicated) and guidance (e.g., providing options and plans to reduce the risk). Risk, together with the two core aspects discussed previously—time and profit—characterize the financial domain well as dimensions of analysis. Time and money are exchangeable resources and risk modulates the efficiency or control of these exchanges. Virtually all tasks in the financial domain involve estimating time-profit-risk relations between entities (stocks, commodities, contracts, currencies, players, one's portfolio), which amounts to understanding the environment, and then maybe performing an action that alters that state (placing a buy or sell order). Since the market is a chaotic system, it is impossible to fully understand the environment and the consequences of one's actions, so all measurements and actions contain some uncertainty.

In our example, the users of the *Profit* tool are investors performing two tasks: building a portfolio and swing trading. In Figure 9 and 10, we provide the respective desktop to do the specific tasks. The goal of building a portfolio is to select promising stable stocks that the investor will hold for a long time before selling. A user building a portfolio performs an exploratory analysis to find good candidate stocks, and the presented desktop is optimized to help him or her find them. On the right-hand view, one can see stocks with larger volume, which can be re-sold more easily in the future, and the ones marked in green present a lower risk since they have a positive valuation. In the bottom left corner, the most popular stocks are displayed, while in the top center field, the user can monitor their portfolio evolution over time.

Swing trading, on the other hand, requires an active effort to identify good moments to buy and sell stocks. This desktop in Figure 10 then shows different information about a stock that helps the trader identify not only good times to buy and sell, but also how much and which stock. Besides the candlestick chart at the bottom enriched with moving average indicators, most of the data displayed is statistics about the order book, where buyers and sellers make their offers, and which is the actual engine behind stock trading. The price graph is an abstraction of what is happening in the book, indicating the average price of the last trades that happened, which is generally a good summary of a stock, but it is far from the complete picture. A user operating this desktop will navigate this interface to gauge potentially good trades, then place orders using the interface in the bottom right corner, and monitor them both on the graph and in the central view.

To successfully perform those two tasks, the investor has to have different types of knowledge $\boxed{K^e}$ (see Section 2.3 for the different knowledge types): domain knowledge about the financial domain, data knowledge, visual encoding and interaction knowledge to understand candlestick charts, treemaps, and line charts, and the different interaction possibilities of the *Profit* tool. If there is a knowledge gap in one of those areas, visualization onboarding (*gain understanding*) and guidance (*gain insight*) can help the user to overcome it ($\boxed{K^e} \rightarrow \boxed{U}$). With the integration of visualization onboarding $\textcircled{O}$ concepts into the *Profit* tool, we can support the users in understanding the domain, data, visualization and interaction. Guidance $\textcircled{G}$ can help the user to choose strategies and support the exploration for building a portfolio.

$$\{\boxed{K^e}, \boxed{D}, \boxed{S}\} \rightarrow \textcircled{O}, \textcircled{G} \rightarrow \{\boxed{Opt}, \boxed{C}, \boxed{Ext}\} \rightarrow \textcircled{P}$$

**Who** can benefit from visualization onboarding and guidance? When we think about the two tasks of building a portfolio and performing swing trading with the *Profit* tool, visualization onboarding can support new users of the tool (investors), but also experienced investors who have never built a portfolio before or have no experience with swing trading. Guidance can support long-term investors and mid- to short-term traders in the respective tasks (data knowledge gap; see Section 2.3). An investigator, such as the financial journalist mentioned above, or anyone just wanting to perform some data analysis on stock data could also be supported by guidance for the first task, as it is largely an exploration task (data & visualization and interaction, in greater detail described in Section 2.3). To sum up, the financial journalist or the investors using the *Profit* tool most probably don't have a knowledge gap in the *domain*, as they are experts in the field. A gap in *data, visualization, and interaction* can occur.

**How** to provide visualization onboarding and guidance? A new user is offered the option of using the tutorial, which is comprised of nine steps. In Figure 14, we present four screenshots highlighting specific parts of the tutorial. *Profit* integrates *visual cues* $\boxed{C}$ (tooltips) with textual descriptions into three parts:

$$\{\boxed{K^e}, \boxed{D}, \boxed{S}\} \rightarrow \textcircled{O} \rightarrow \{\boxed{C}\} \rightarrow \textcircled{P}$$

These three parts are: learning how to take a position by configuring and placing orders (steps 1 to 4, Fig. 14a), learning how to monitor and close ones position (steps 5 and 6, Fig. 14b), and learning how to explore more options and get support (steps 7 to 9, Figs. 14c, 14d). In the last step (9, Fig. 14d), the user is directed towards more in-software onboarding tutorials (Visualization Onboarding /Onb), or to additional study material available in multiple media.

$$\{\boxed{K^e}, \boxed{D}, \boxed{S}\} \rightarrow \textcircled{O} \rightarrow \{\boxed{Ext}\} \rightarrow \textcircled{P}$$

In the YouTube channel $\boxed{Ext}$, for instance, content is published daily with guidance on how to use the software for different purposes. These onboarding concepts can be used to support the investor in building a portfolio as well as in performing swing trading.

Guidance $\textcircled{G}$ is not explicitly provided in the *Profit* tool. However, various supporting mechanisms support the user in building a portfolio or performing swing trading. In the light of our framework, this can be considered as guidance. First, *Profit* provides different desktop configurations the user can save, load, and share $\boxed{Opt}$. As exemplified in Figure 9 and Figure 10, this can be used to guide users towards suitable setups for performing certain tasks. Furthermore, most components and indicators are hyper-parametrized, and this parametrization is task-sensitive. Providing preset parameters for certain strategies and tasks is a weak form of guidance, which could be improved by dynamic, user-centered recommendations. A user who has never used a certain indicator, but is performing swing trading, for instance, might identify this preset from the list and have a better chance at gaining insights for this task.

$$\{\boxed{K^e}, \boxed{D}, \boxed{S}\} \rightarrow \textcircled{G} \rightarrow \{\boxed{Opt}\} \rightarrow \textcircled{P}$$

Finally, *Profit* provides a simulator that allows the user to replay the behavior of a stock at a particular time, and test strategies. While this is not explicitly visualization onboarding or guidance, it provides support for both, as one can gain both understanding and insights from the simulation. Testing a strategy on past data is strongly recommended, and unexpected results and patterns can be uncovered.

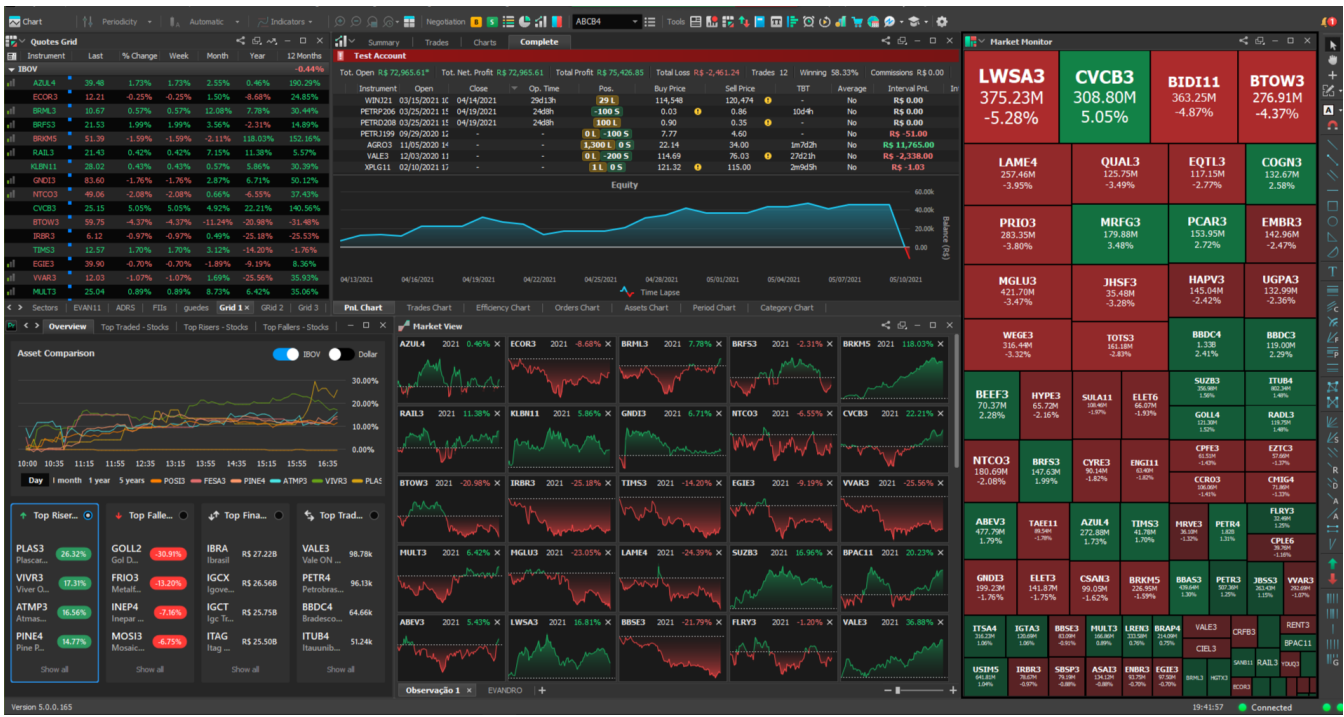

Fig. 9. The adapted desktop for performing the task of building a portfolio which we present as more in line with FA. Most of the screen space is used to visualize stocks to identify good candidates.

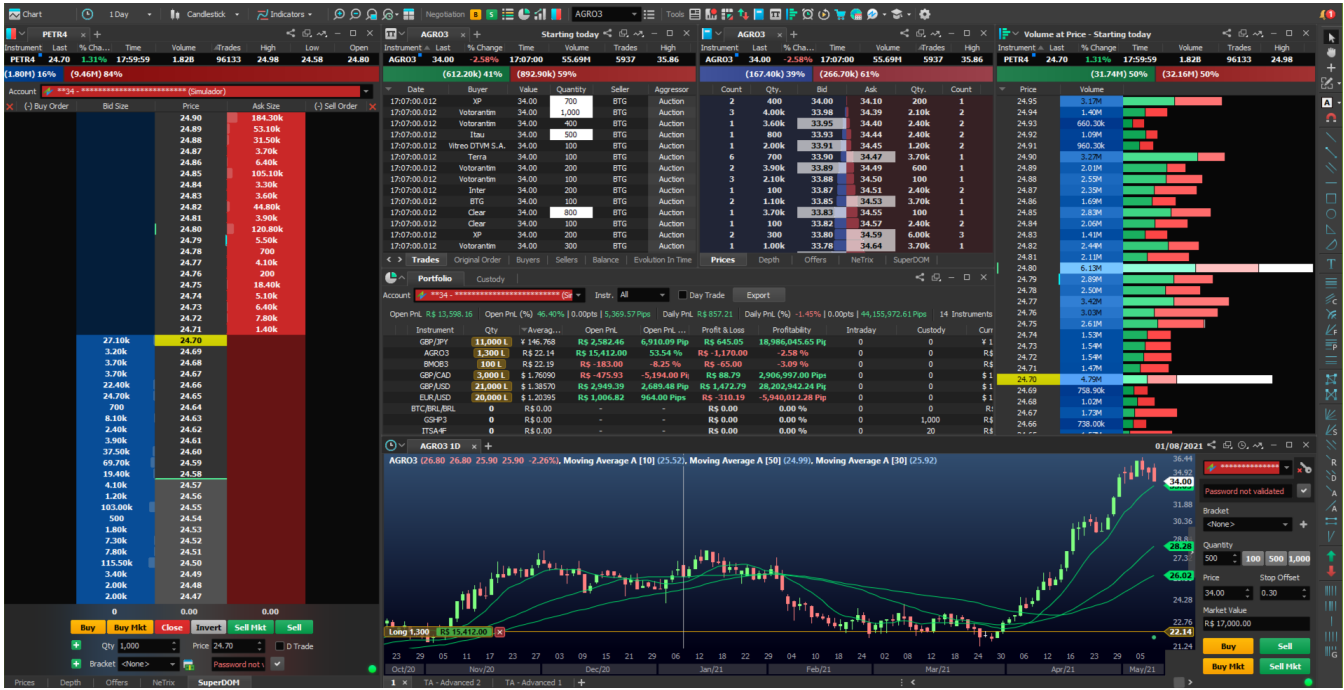

Fig. 10. Swing trading desktops configuration, which is rather associated with TA tasks. In this setup, the screen shows different information about the stocks to find good times to buy and sell stocks to which price.

**When** to provide guidance and/or onboarding? Visualization onboarding and guidance are required at different points in the data exploration with *Profit*. For the task of building a portfolio, visualization onboarding can support first-time users or during use when new features are activated. Guidance can be used to choose the right parameters for indexes and encodings, so that the user can easily search for patterns. For the task of *performing swing trading*, visualization onboarding of first-time users is even more important, as there is more risk involved. Additionally, when some real-time event occurs, visualization onboarding can be called up. Guidance, on the other hand, is provided mainly at two times: before operation, for example while selecting an interesting stock or financial instrument to operate; and during operation, actively assisting the trading process by showing potential entry and exit points for trades.

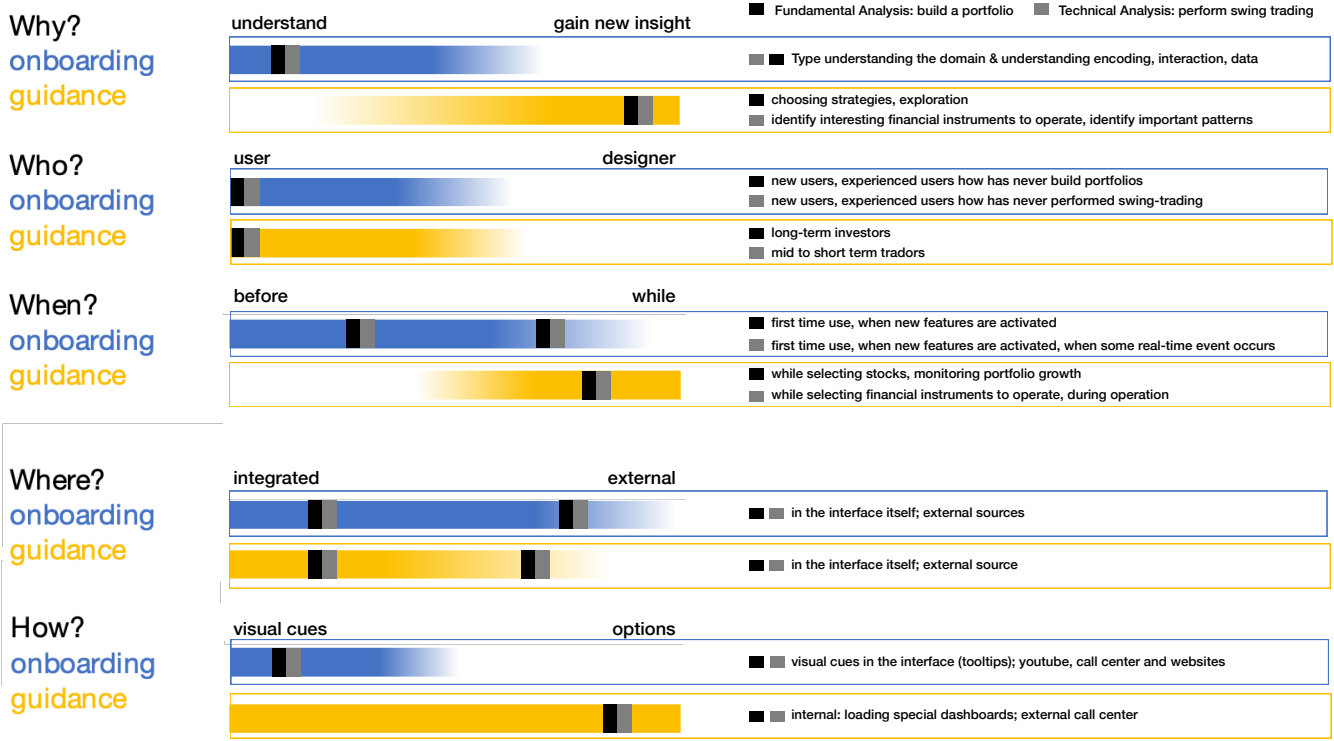

Fig. 11. Figure showing the comparison of two high-level tasks in the context of building a portfolio (Fundamental Analysis (FA)) and performing swing trading (Technical Analysis task (TA)) in the *Profit* VA tool, respectively, according to the criteria described in Fig. 2. Exemplarily, for the question of **When** to provide guidance and/or onboarding? guidance and onboarding can be provided differently. For example, guidance is integrated in the tool providing user assistance **while** using the tool, contrary, onboarding also supporting users **before** actually starting the data exploration with the *Profit*. In the light of the using the *Profit* tool for building a **portfolio**/**performing swing trading** the integrated **onboarding** concepts can support the user when opening the *Profit* tool for the first time, or when some real-time events occurs, and/or when new features are activated. Besides, **guidance** supports users **while** selecting stocks, monitoring portfolio growth (FA), and also while selecting financial instruments to operate.

**Where** to provide visualization onboarding and guidance? For visualization onboarding, the *Profit* tool integrates a nine-step onboarding process within the main interface. Then, in the last step of these in-place onboarding instructions, the system points to additional study material which is available in multiple media, such as additional in-software tutorials, social media, and an integrated support hotline. In the YouTube channel, for instance, content is published daily with support on how to use the software for different strategies. This can be helpful while performing both tasks of building a portfolio and swing trading. Most features also have a help component, which at first sight can be considered pure onboarding, but can provide guidance when linked to additional support content. For instance, if the user has never used Renko charts, and clicking the help button on a Renko chart of the XYZ index suggests an introductory YouTube video on this topic, this qualifies as guidance.

In summary, there are many ways in which our framework can be applied to make use of visualization onboarding and guidance whenever a user has to interact with data to perform a task.

## 6 DISCUSSION & LESSONS LEARNED

In the following, we present lessons learned and limitations of our descriptive model.

### 6.1 Integration of Guidance & Visualization Onboarding

Our descriptive model shows theoretically how visualization onboarding and guidance can be integrated into the VA process and how their different components interact with each other (see Section 3). In terms of current approaches to visualization onboarding, we can list some solutions of commercial VA tools, such as [34, 50, 56, 57, 63, 66]. When it comes to guidance, some approaches can also be found in the literature [13, 42–44]. However, hardly any VA system integrates both concepts. Therefore, we present how visualization onboarding and guidance can be integrated into a VA system to support users in the financial domain based on a usage scenario with the VA tool *Profit*. In Section 5, we describe the use of visualization onboarding and guidance along the five questions of our framework. We decided on the financial domain as it is complex and hard to fully understand in detail. For novices in the financial domain (e.g., in financial analysis or a journalist for a newspaper), basic concepts, terms and visual encoding (e.g., sticks) have to be explained to successfully perform financial analysis.

Lessons Learned from the Usage Scenario: When applying the Why/Who/When/Where/How analysis to both tasks (Fig. 11) within *Profit*, we can outline the differences and similarities between them, but also between the visualization onboarding and guidance dimensions. On the "Why" criterion, for instance, visualization onboarding and guidance are diametrically opposed on the understanding/insight axis, but both tasks share the same motivation for visualization onboarding. The "Who", however, has both visualization onboarding and guidance matching on the user/designer axis, but the user profiles are different. Finally, the best way to distinguish both is in the "How", which can be characterized well in the financial domain for the active operational aspect (also linked to "When"), even if it has the idiosyncrasy of providing human-based visualization onboarding and guidance through an external call center. In summary, there are many ways that our framework can be applied to make use of **guidance** and **visualization onboarding** whenever a user has to engage with data to perform a task.

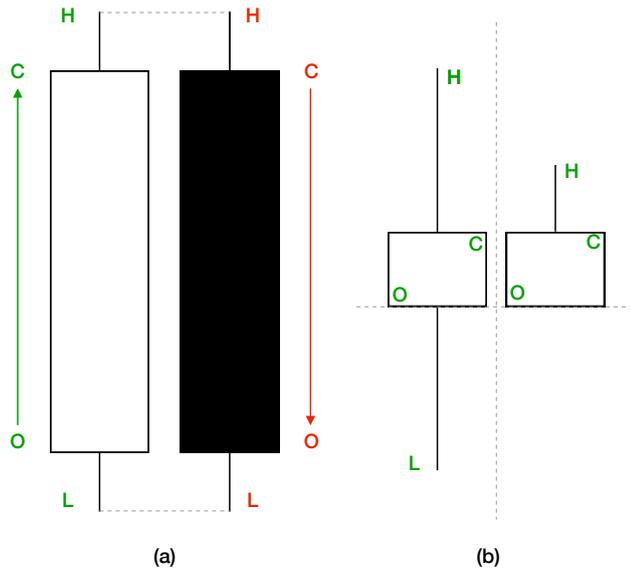

Fig. 12. (a) Basic anatomy of a candlestick chart with all its essential components: opening price (O), closing price (C), highest (H), and lowest (L). The white or green color is associated with a rise in price ($C > O$), and a red or black color is used for a fall in price ($C < O$). The painted area between O and C is the "real" body of the candlestick, and the lines that stretch to H and L are its shadow or tail. (b) Comparison between two visually similar patterns, which have the same real body but highly distinct shadows, and therefore represent different stories. Both candles show the same positive price variation ($C > O$) but on the right, the price oscillation was much smaller than on the left, indicating that market sentiment is more stable on how to price this stock during the period.

### 6.2 Descriptive Model of Visualization Onboarding and Guidance

According to Beaudouin-Lafon [5], models can exhibit three types of characteristics: (1) *descriptive*: the ability to describe a wide range of existing methods; (2) *evaluative*: enable the assessment of multiple design alternatives, and (3) *generative*: help in designing new methods. Our proposed model is of the descriptive kind (see Section 4), as it has the ability to systematically describe and present relevant aspects of visualization onboarding and guidance.

Lessons Learned from building the Descriptive Model: Thus, our model helps visualization designers who can use it as a blueprint for future VA tools integrating visualization onboarding and guidance components. First of all, they can decide on how and when to integrate visualization onboarding or/and guidance into their VA system (see Figure 2). On the machine side of the descriptive model (Figure 1), the designer can see that the data, for example, can serve as an input for the planned onboarding. Furthermore, s/he can decide on how to provide the onboarding perceived by the user through integrating visual cues, for example.

### 6.3 Limitations

One limitation of the model is that it does not provide a step-by-step architecture (e.g., design patterns, algorithms, data structures) or concrete design guidelines for visualization onboarding and guidance concepts (e.g., instructional material in the form of videos, texts, etc., highlighting concepts for leading the focus to interesting elements in the VA system). Since this model is a high-level blueprint, another limitation is the possible depth of describing the user's cognitive processes, perception, and tacit knowledge generation. Moreover, additional research needs to be carried out to extend the model beyond its descriptive nature

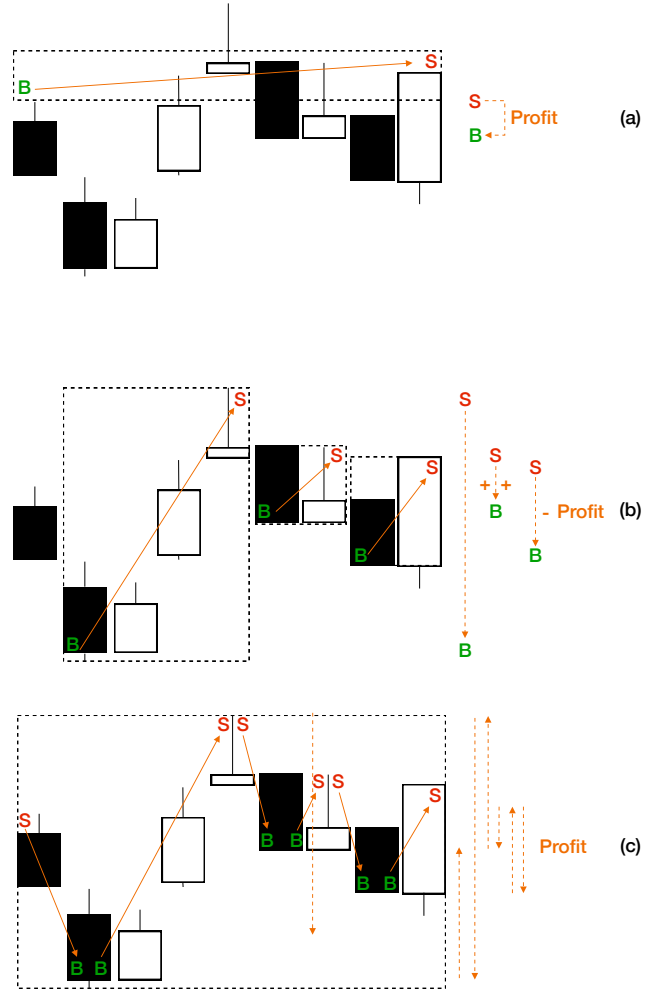

Fig. 13. Distinct trading scenarios with the same candlestick patterns but different strategies. (a) A stock is bought and held for a long time, then sold. (b) Multiple trades are performed in the same period, taking advantage of low and high price variations to optimize profit. (c) If shorting is allowed, all movements can be turned into profit, but at a higher risk. The rectangles encapsulate the space where each position is open, and can be cues to understand the risk in each situation. Profit can be determined by the difference in price between entering and exiting operations, the orange arrows, which is essentially their vertical component, indicated on the right by the dotted arrows.

to a more generative and evaluative one for visualization onboarding and guidance.

## 7 CONCLUSION AND FUTURE PERSPECTIVES

In this paper, we contributed a descriptive model of visualization onboarding and guidance in VA. When it comes to such models, we assess their differences and similarities along the questions of Why/Who/When/Where/How, thereby leading to their effective and thoughtful use in VA. We also show an application of the model in the financial domain illustrating the importance of visualization onboarding and guidance for an effective analysis.

The in-depth discussions and comparisons opened up several future perspectives: (1) Effective assistance in the form of guidance and/or onboarding during the visual exploration process needs to be evaluated in more detail with a particular focus on real-world and real-time scenarios. This calls for appropriate evaluation metrics and processes, which are open challenges in VA [31]; (2) Visualization onboarding and guidance concepts require flexible and adaptive approaches according to the prior knowledge of the users, which could be determined by the knowledge gap, the kinds of exploration, as well as the interaction behavior of the users; (3) Tacit and explicit knowledge are crucial for visualization onboarding and guidance. Consequently, the reliability of such knowledge needs to be assessed.

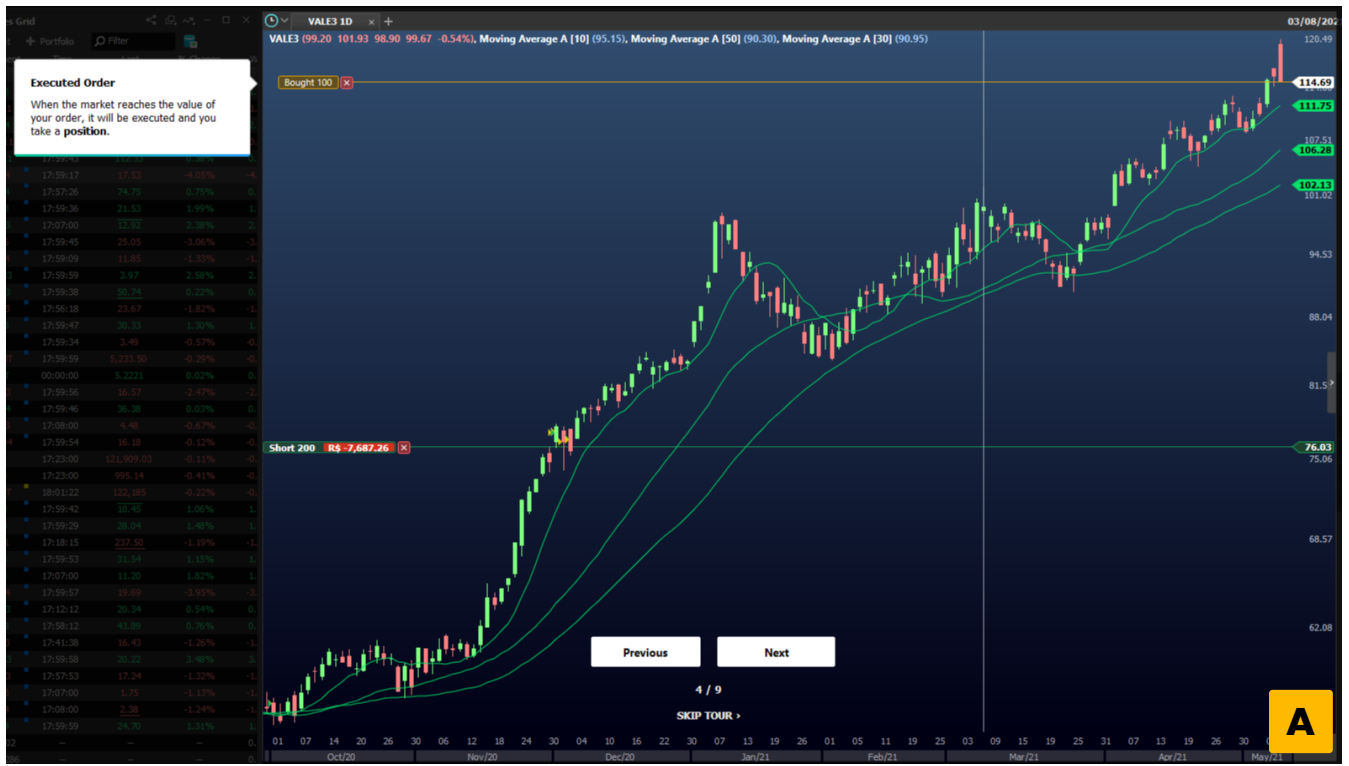
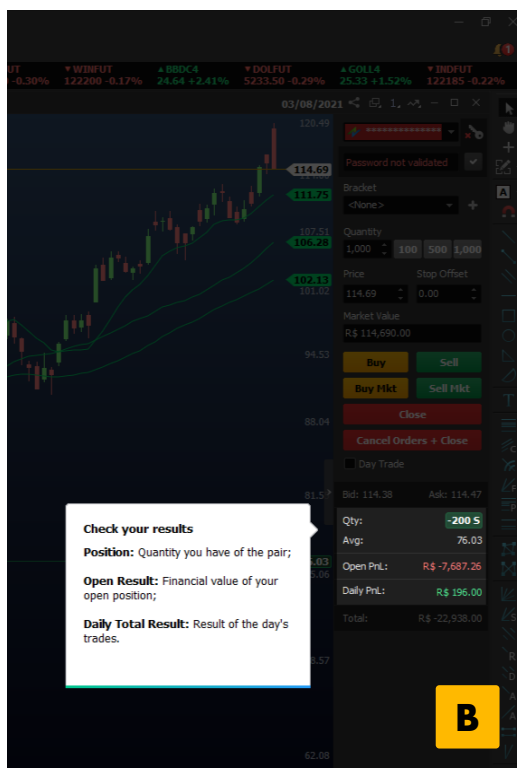
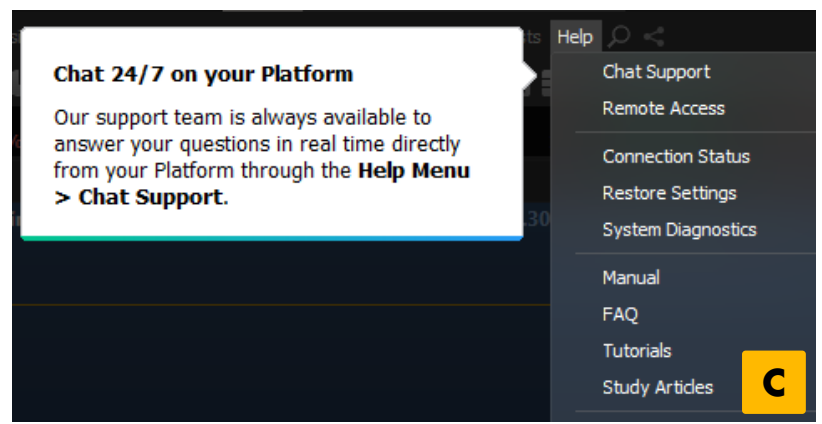
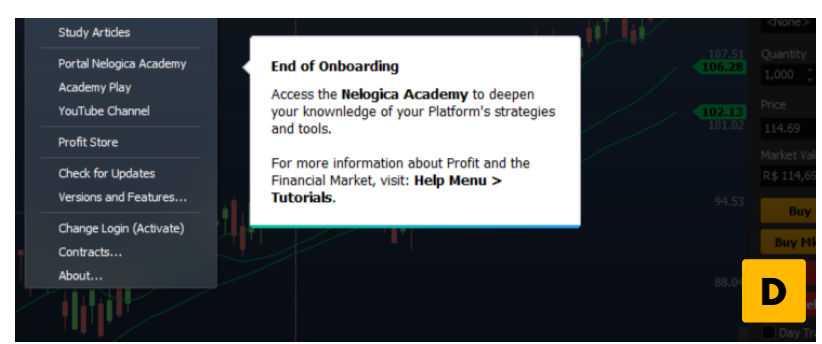

Fig. 14. Screenshots of *Profit's* first-time user onboarding, which is comprised of nine steps. Here we show four out of nine steps: **A** Overview of the interface, the main graph and how to place orders. **B** Monitoring trading results. **C** In the financial domain, extreme importance is attached to live support, which is why it is presented as an essential tool during onboarding. **D** Further options, both internal and external to the software are provided to proceed with the learning curve. The images are all in screen resolution to reproduce the users' experience. They should be zoomed in on, and read left to right, top to bottom.

**ACKNOWLEDGMENTS**

We would like to thank Frederico Limberger and Nelogica for their support in evaluating Profit, and all participants who attended our Application Spotlight Workshop at the IEEE Vis 2019 discussing and providing very helpful feedback to our approach. This work was supported by the Austrian Science Fund (FWF) as part of the projects VisOnFire and KnoVA (#P27975-NBL, #P31419-N31), the Vienna Science and Technology Fund (WWTF) via the grant ICT19-047 (GuidedVA), the Austrian Ministry for Transport, Innovation and Technology (BMVIT) under the ICT of the Future program via the SEVA project (#874018), as well as by the FFG, Contract No. 854184: "Pro2Future" is funded within the Austrian COMET Program Competence Centers for Excellent Technologies under the auspices of the Austrian Federal Ministry for Transport, Innovation and Technology, the Austrian Federal Ministry for Digital and Economic Affairs, and of the Provinces of Upper Austria and Styria. COMET is managed by the Austrian Research Promotion Agency FFG.